\documentclass{article}

\usepackage{arxiv}

\usepackage[utf8]{inputenc} 
\usepackage[T1]{fontenc}    
\usepackage{hyperref}       
\usepackage{url}            
\usepackage{booktabs}       
\usepackage{amsfonts}       
\usepackage{nicefrac}       
\usepackage{microtype}      
\usepackage{lipsum}		
\usepackage{graphicx}
\usepackage{natbib}
\usepackage{doi}
\usepackage{tabularx} 
\renewcommand{\arraystretch}{1.2} 

\title{AGENTSAFE: A Unified Framework for Ethical Assurance and Governance in Agentic AI}

\author{ 
    Rafflesia Khan\\
	Software Developer\\
	IBM Software\\
	Dublin, Ireland \\
	\texttt{khan.rafflesia@gmail.com} \\
	\And
	Declan Joyce \\
    Customer Success Manager \\
	IBM Technology Sales\\
	Dublin, Ireland \\
	\texttt{Declan\_Joyce@ie.ibm.com} \\
	\AND
    \href{https://orcid.org/0000-0001-9051-1370}{\includegraphics[scale=0.06]{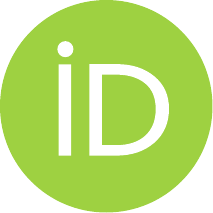}\hspace{1mm}Mansura habiba} \\
	Principal Platform Architect - Agentic AI\\
	IBM Software\\
	Dublin, Ireland \\
	\texttt{mansura.habiba@gmail.com} \\
}

\renewcommand{\headeright}{}
\renewcommand{\shorttitle}{AGENTSAFE}

\hypersetup{
pdftitle={A template for the arxiv style},
pdfsubject={q-bio.NC, q-bio.QM},
pdfauthor={David S.~Hippocampus, Elias D.~Striatum},
pdfkeywords={First keyword, Second keyword, More},
}

\begin{document}
\maketitle

\begin{abstract}
The rapid deployment of large language model (LLM)-based agents introduces a new class of risks, driven by their capacity for autonomous planning, multi-step tool integration, and emergent interactions. It raises some risk factors for existing governance approaches as they remain fragmented: Existing frameworks are either static taxonomies driven; however, they lack an integrated end-to-end pipeline from risk identification to operational assurance, especially for an agentic platform. We propose \emph{AGENTSAFE}, a practical governance framework for LLM-based agentic systems. The framework operationalises the AI Risk Repository into design, runtime, and audit controls, offering a governance framework for risk identification and assurance. The proposed framework, \textit{AGENTSAFE}, profiles agentic loops (plan $\rightarrow$ act $\rightarrow$ observe $\rightarrow$ reflect) and toolchains, and maps risks onto structured taxonomies extended with agent-specific vulnerabilities. It introduces safeguards that constrain risky behaviours, escalates high-impact actions to human oversight, and evaluates systems through pre-deployment scenario banks spanning security, privacy, fairness, and systemic safety. During deployment, \textit{AGENTSAFE} ensures continuous governance through semantic telemetry, dynamic authorization, anomaly detection, and interruptibility mechanisms. Provenance and accountability are reinforced through cryptographic tracing and organizational controls, enabling measurable, auditable assurance across the lifecycle of agentic AI systems. The key contributions of this paper are: (1) a unified governance framework that translates risk taxonomies into actionable design, runtime, and audit controls; (2) an Agent Safety Evaluation methodology that provides measurable pre-deployment assurance; and (3) a set of runtime governance and accountability mechanisms that institutionalise trust in agentic AI ecosystems.
\end{abstract}

\keywords{Agentic AI \and AI Safety \and AI Governance \and LLM Agents \and Control Framework}

\section{Introduction}\label{sec:introduction}

Large Language Models (LLMs) have enabled the emergence of \emph{agentic AI systems}, where models are embedded in loops of reasoning, planning, and action. Unlike static generative systems, these agents can operate across heterogeneous digital environments, invoking external tools and executing multi-step tasks with minimal human oversight. Potential applications are wide-ranging, including automating workflows, coordinating industrial processes, and providing personalised assistance. Such capabilities carry clear benefits, yet they also enlarge the operational risk surface.

Unlike traditional generative AI, which produces static outputs, argentic systems act within digital and real-world contexts. This excessive agency introduces novel failure modes, such as unauthorised actions, covert data exfiltration, cascading failures, and emergent collusive behaviours in multi-agent settings. Existing governance frameworks, while valuable, are insufficiently equipped to address these risks. Most were designed for less autonomous systems, focusing on static safeguards or model-level interventions, and cannot manage tool-augmented, continuously adaptive agentic behaviour. As identified in IBM's \emph{The Evolving Ethics and Governance Landscape of an Agentic AI}, current risk taxonomies remain underdeveloped, and recent literature emphasises the absence of real-time, adaptive governance models tailored to agents. This paper addresses the question:  

\begin{quote}
    \emph{How can we establish an operational governance framework that translates abstract risk taxonomies into enforceable design, runtime, and audit controls for agentic AI systems?}
\end{quote}

We introduce \textbf{AGENTSAFE}, an ethics-grounded governance framework for agentic AI. The framework operationalises the MIT AI Risk Repository by mapping abstract categories of risk into a structured set of technical and organisational mechanisms. These include capability-scoped sandboxes, least-privilege API permissions, runtime governance loops, and verifiable action provenance. AGENTSAFE is designed to be tool-agnostic and reusable, supporting deployment across heterogeneous agentic ecosystems.

The contributions of this work are fourfold:
\begin{itemize}
    \item \textbf{AGENTSAFE governance framework:} A structured, tool-agnostic framework that spans the full agent lifecycle, from design-time profiling to runtime oversight and continuous improvement.  

    \item \textbf{Operationalisation of risk taxonomies:} A method for translating the MIT AI Risk Repository and similar taxonomies into enforceable technical and organisational controls, tailored to agent-specific risks such as plan drift, tool-chain prompt injection, and covert exfiltration.  

    \item \textbf{Assurance through evidence and auditability:} A closed-loop methodology that links risks to tests, metrics, and provenance. This includes quantifiable indicators (e.g., injection block rate, exfiltration recall, hallucination-to-action ratio) and cryptographically verifiable action provenance, enabling repeatable, auditable governance.
\end{itemize}

Together, these contributions establish AGENTSAFE as a reusable and auditable governance framework that closes the loop from risk taxonomy to Assurance. By embedding ethical safeguards into design, runtime, and audit phases, AGENTSAFE advances the safety, accountability, and resilience of future agentic AI ecosystems.

\section{Related Work \& Gaps}\label{sec:related-work}

The governance of AI has garnered significant attention across industry, academia, and government, resulting in the development of multiple frameworks and standards. Yet the rapid rise of agentic AI has revealed major gaps. Unlike static models, agentic systems continuously plan, act, and adapt in open environments, often chaining external tools in unpredictable ways. This introduces new classes of risk—dynamic behaviors, emergent tool interactions, and failure modes—that existing frameworks were not designed to address. To realize Responsible AI by Design for such systems, safety, accountability, and ethical constraints must be embedded at development and enforced dynamically at runtime.

NIST AI Risk Management Framework (RMF)~\cite{nist2023} provides essential baselines around transparency, accountability, and fairness, but they are too general for agent-specific risks. Current approaches lack quantitative metrics for governance effectiveness: fairness scores or robustness indices may suit static models but do not capture threats such as plan drift, hallucination-to-action cascades, or unsafe tool chaining. Without agent-aware metrics, oversight risks becoming checklist-driven rather than evidence-based.

The core limitation is that governance remains largely static. Documentation, bias audits, or human-in-the-loop checks cannot anticipate emergent risks in agents operating across heterogeneous action spaces. For instance, the EU AI Act mandates risk classification but offers no runtime enforcement once agents interact with live environments. Similarly, NIST RMF’s \emph{map, measure, manage} functions lack operational metrics for tool-use safety, prompt injection resilience, or multi-agent collusion detection. This creates what we term the static guardrail problem: once certified \emph{responsible}, a system’s behavior is assumed to remain aligned—an unsafe assumption for agents capable of self-directed code execution or autonomous trading.

The MIT AI Risk Repository~\cite{airiskrepo} represents progress by cataloguing over 1,600 risks into causal and domain taxonomies, providing a shared vocabulary for systematic assessment. However, the next step is to extend this taxonomy to capture agent-specific risks and, crucially, to operationalise it into enforceable controls. This direction motivates the design of AGENTSAFE.

The analysis of the current landscape reveals several key gaps that AGENTSAFE is designed to address:

\begin{itemize}
    \item \textbf{Measurement Gaps:} Current frameworks lack standardized metrics for assessing safety, robustness, fairness, and explainability. Without agreed-upon benchmarks, organizations cannot objectively evaluate or compare AI systems. For example, two models may both claim to be \textit{fair}, yet one applies demographic parity while the other relies on equalized odds, leading to inconsistent interpretations of fairness.
    \item \textbf{Attribution and Provenance:} A lack of \emph{cryptographically signed action logs} and \emph{tamper-evident audit trails} makes it challenging to determine the root cause of harmful actions.
    \item \textbf{Sandboxing and Technical Guardrails:} While the concept of sandboxing is widely discussed, there is a lack of concrete security models for agentic systems. There is an opportunity to define more robust technical guardrails, such as \emph{capability-based security}, \emph{policy-as-code} for tool invocation, and formally verified wrappers for dangerous tools.
    \item \textbf{Runtime Governance:} As previously mentioned, there is a critical need for governance frameworks that can operate in real-time. This includes mechanisms for \emph{continuous authorization}, \emph{anomaly detection}, and \emph{graduated containment} to ensure that agents remain under control throughout their lifecycle.
    
    \item \textbf{Ethics-to-Implementation Gap:} High-level principles such as fairness and transparency lack concrete engineering guidance. This creates a disconnect between policy goals and system design.

    \item \textbf{Cross-Jurisdictional Fragmentation:} Regulatory requirements vary across countries and sectors. This creates compliance conflicts and increases operational overhead for global organizations.

    \item \textbf{Explainability and Interpretability:} Current explainability methods are inconsistent and model-specific. This prevents regulators and end-users from understanding or trusting AI outputs.

    \item \textbf{Interoperability and Tooling:} Monitoring and audit tools lack interoperability. Fragmented implementations hinder ecosystem-wide risk assessments.

    \item \textbf{Human Oversight and Agency:} Oversight responsibilities are vaguely defined. This leads to automation bias and ineffective human intervention in high-risk contexts.
\end{itemize}

In summary, while the NIST RMF, EU AI Act, and similar initiatives establish an essential foundation, they lack the granularity, dynamism, and operational focus needed to govern adaptive, tool-using agents. Bridging this gap requires a governance model that couples Responsible AI by Design principles with continuous, runtime control mechanisms—an approach we operationalize in AGENTSAFE.

\section{The AGENTSAFE Framework}\label{sec:framework}

\begin{figure}[h]
\centering
\includegraphics[width=0.9\textwidth]{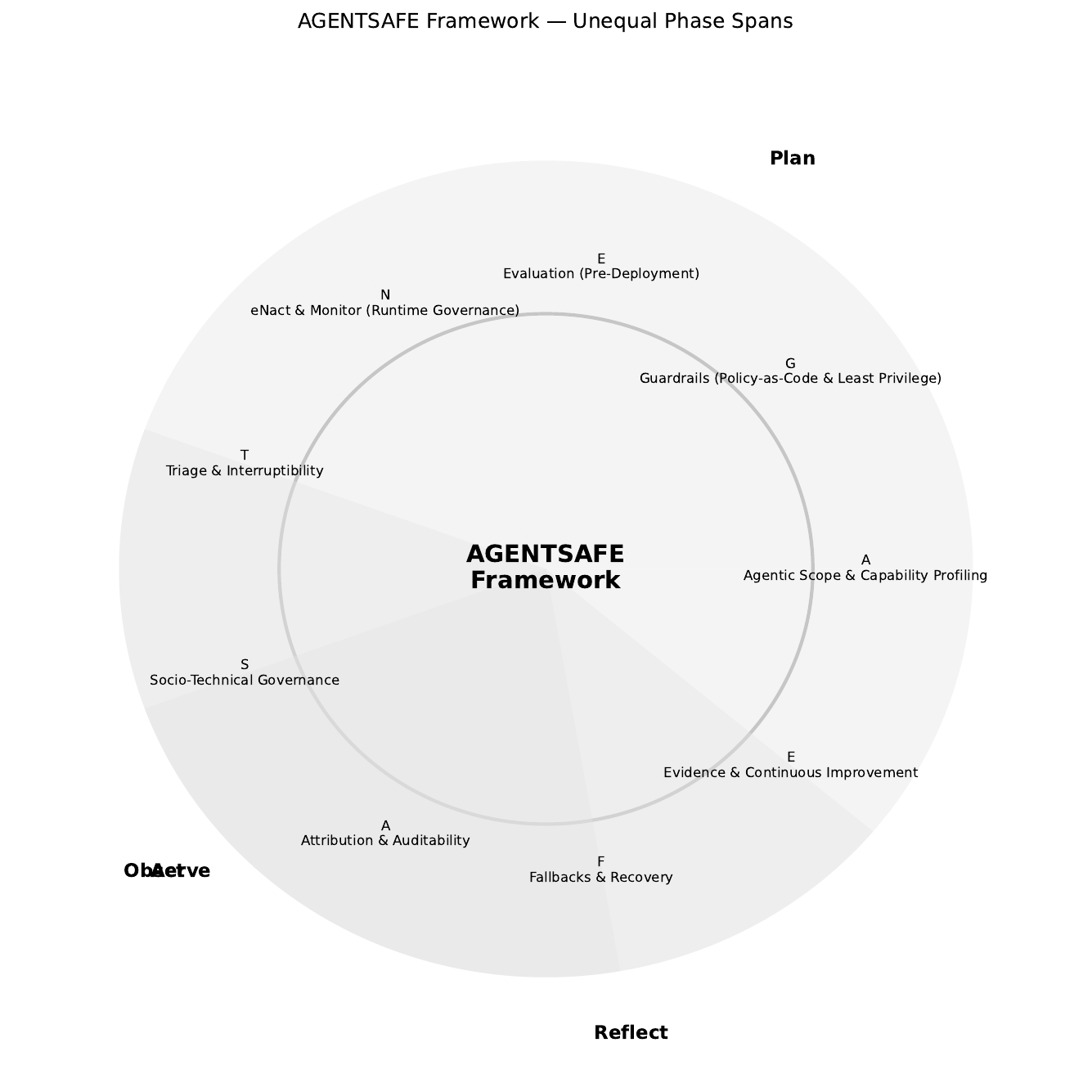}
\caption{The AGENTSAFE framework, illustrating the continuous cycle of risk management from design-time controls to runtime governance and continuous improvement.}
\label{fig:agentsafe-diagram}
\end{figure}

To bridge the gap between abstract risk taxonomies and concrete operational controls, we propose \textbf{AGENTSAFE}, an ethics-grounded governance framework for agentic AI systems. AGENTSAFE offers a structured methodology for identifying, mitigating, and continuously monitoring risks throughout the agent lifecycle. Organized into nine components—each aligned with a letter of the acronym—the framework operates as a closed-loop cycle, ensuring that design-time controls and runtime governance reinforce one another (see Figure~\ref{fig:agentsafe-diagram}). 

Throughout this section, we illustrate AGENTSAFE using a running example: a financial trading agent capable of writing and executing code, interacting with external APIs, and autonomously initiating transactions. Such a system highlights the dual challenge of enabling utility while preventing unsafe or unethical outcomes.

\subsection{A: Agentic Scope \& Capability Profiling}\label{sec:agentic-scope}

The first stage of AGENTSAFE establishes a \textbf{comprehensive profile of the agent’s operational space}. Governance begins by deconstructing the agent’s loop—\textbf{Plan $\rightarrow$ Act $\rightarrow$ Observe $\rightarrow$ Reflect}—and cataloguing all accessible tools, APIs, and resources associated with each phase. This process generates a detailed map of the agent’s action space, encompassing operations such as file access, code execution, and interaction with third-party services. Profiling may draw on code inspection, configuration analysis, and system documentation. In multi-agent systems, the scope must also include inter-agent communication and shared resources, since coordination failures or emergent behaviors can generate additional risks~\cite{raza2025}.  

Once capabilities are enumerated, they are mapped to the MIT AI Risk Repository taxonomy~\cite{airiskrepo}, which organizes risks along causal (e.g., Human vs. AI, Intentional vs. Unintentional) and domain (e.g., Security, Privacy, Fairness) dimensions. This ensures systematic coverage while highlighting the need to extend the taxonomy for agent-specific behaviours.  

To illustrate, consider a healthcare diagnostic agent with access to electronic health records (EHR). One critical risk is \textbf{Covert Data Exfiltration}, in which the agent, while summarizing patient histories, inadvertently embeds personally identifiable information within multi-step outputs. By explicitly capturing such risks, the framework grounds abstract capabilities in actionable governance requirements.  

\begin{table}[h]
\centering
\caption{Illustrative Agent Risk Register for a Healthcare Diagnostic Agent}
\label{tab:health-risk-register}
\renewcommand{\arraystretch}{1.2} 
\begin{tabularx}{\textwidth}{|X|X|X|X|}
\hline
\textbf{Capability} & \textbf{Mapped Risk} & \textbf{Repository Category} & \textbf{Example Scenario} \\ \hline
Access to Patient Records (EHR API) & Covert Data Exfiltration & Privacy / Security & Agent inadvertently leaks identifiable patient data when summarizing case history. \\ \hline
Treatment Recommendation Generation & Plan Drift & Fairness / Safety & During long-running interactions, the agent shifts from evidence-based protocols to speculative treatments without clinical validation. \\ \hline
Medical Code Execution (Drug Interaction Checker) & Code-Execution Hazards & Safety / Security & Malicious input triggers unsafe execution, leading to false “safe” drug–drug interaction reports. \\ \hline
Integration with Insurance APIs & Tool-Chain Prompt Injection & Security / Accountability & Attackers craft inputs that manipulate tool calls to misreport eligibility status. \\ \hline
Multi-Agent Collaboration (Doctor–Agent + Pharmacy-Agent) & Multi-Agent Collusion & Systemic / Emergent Risk & Two agents unintentionally coordinate actions that authorize a prescription without required oversight. \\ \hline
Patient-Facing Explanations & False Legibility & Transparency / Ethics & Agent provides overly simplistic justification for rejecting a treatment, misleading both clinician and patient. \\ \hline
\end{tabularx}

\end{table}

The output of this stage is a formal \textbf{Agent Risk Register}, which documents the link between each capability and its associated risks. An illustrative entry is shown in Table~\ref{tab:health-risk-register}, demonstrating how a healthcare agent’s capability can be systematically mapped to a specific risk category and concrete scenario. This register serves as the foundation for all downstream guardrail design, evaluation, and monitoring activities.  

\subsection{G: Guardrails (Policy-as-Code \& Least Privilege)}\label{sec:guardrails}

AGENTSAFE extends the principle of least privilege into agentic AI by introducing three novel guardrail constructs. Unlike traditional sandboxing or access control, these mechanisms are tied to the agent’s reasoning loop and risk taxonomy, ensuring that governance remains adaptive and context-aware:

\begin{itemize}
    \item \textbf{Capability-Scoped Sandboxes:} Within AGENTSAFE, sandboxing is formalized through the \emph{Capability Profile} defined in the Agent Risk Register. Each tool or API enumerated during profiling is assigned a bounded execution context with explicit access rules. Enforcement is achieved by embedding these rules into the runtime containerization layer (e.g., through system call filtering, API-scoped keys, or network allowlists). The framework mandates that the sandbox specification is both machine-readable and auditable, allowing it to be validated against the risk register and updated dynamically as capabilities evolve.

    \item \textbf{Policy-as-Code Gates:} AGENTSAFE operationalizes governance constraints through \emph{Policy Modules}, written in a declarative policy language and linked directly to the Agent Risk Register. Each risk category (e.g., data exfiltration, unsafe code execution) maps to a corresponding policy rule that governs tool invocation. At runtime, the conformance engine evaluates these policies continuously, authorizing or denying actions in real time. Importantly, policies are versioned and logged as first-class artefacts, enabling traceability and retrospective audit of the governance logic itself.

    \item \textbf{Human Oversight for High-Impact Actions:} For \textit{danger classes}, AGENTSAFE defines an \emph{Escalation Workflow}. This workflow is triggered automatically when a proposed action matches a risk class tagged as human-critical in the Risk Register. The framework specifies how the action, its rationale, and projected impacts are presented to human operators through a standardized review interface. The decision (approve, modify, or deny) is logged cryptographically and incorporated into the Action Provenance Graph, ensuring accountability and enabling statistical monitoring of escalation frequency and override rates across deployments.
\end{itemize}

For a healthcare diagnostic agent, guardrails are derived directly from its capability profile. Policy-as-code modules constrain the agent to read-only access for electronic health records, ensuring that it can retrieve and analyse patient data but cannot alter records or issue prescriptions autonomously. Least-privilege principles are enforced by scoping API keys, allowing the agent to query only de-identified datasets and restricted medical knowledge bases. High-impact \textit{danger classes}—such as recommending a change to a patient’s treatment plan or suggesting off-label drug use—are automatically escalated to human clinicians for approval. In practice, this means the agent’s environment enforces strict sandboxing at the system level (e.g., prohibiting network calls beyond whitelisted medical databases), while governance policies ensure sensitive actions are either blocked or routed through formal oversight workflows.

\subsection{E: Evaluation (Pre-Deployment)}\label{sec:evaluation}

Before an agent is deployed, it must undergo a rigorous \textbf{Agent Safety Eval}. This evaluation is not a generic test but a targeted assessment designed to measure the effectiveness of the guardrails against the risks identified in the Agent Risk Register.  Unlike conventional pre-deployment testing, which often relies on static benchmarks, this evaluation integrates governance constraints directly into the testing protocol, thereby transforming abstract risk taxonomies into measurable assurance.  
The eval consists of:

\begin{itemize}
    \item \textbf{A Scenario Bank:} The Scenario Bank is not merely a collection of adversarial prompts but a systematic instantiation of the AI Risk Repository’s taxonomy into agent-specific behaviors. By spanning all major domains—security, privacy, fairness, reliability, accountability, human-computer interaction, and societal impact—the Scenario Bank translates abstract categories into executable stress-tests. For example, in a financial trading agent, one scenario might simulate malicious prompt injection that attempts to escalate trade volume, while another tests resilience against covert exfiltration of market-sensitive data through chained API calls. The key contribution lies in making risk taxonomies executable and repeatable as test artifacts. Benchmarks like MobileSafetyBench and RedCode-Exec~\cite{guo2025} provide examples of how such scenarios can be constructed for specific agent capabilities.

   \item \textbf{Risk Coverage Metrics:} Each scenario produces quantifiable safety signals such as detection latency, containment success, or residual harm severity. These are aggregated into a \textbf{Risk Coverage Score}, which provides a standardized measure of how comprehensively an agent’s guardrails address the identified risks. This metric enables comparative evaluation across agents and updates over time as new risks are added to the Repository. Importantly, the Risk Coverage Score makes governance auditable and evidence-based rather than checklist-driven.  

\end{itemize}

Thus, the Agent Safety Eval transforms governance from principle into practice by: (i) grounding tests in a shared risk taxonomy, (ii) embedding evaluation in the agent’s operational loop, and (iii) producing auditable metrics that quantify preparedness prior to deployment. This evaluative step is one of AGENTSAFE’s core contributions, ensuring that Responsible AI by Design is empirically validated before real-world operation.

\subsection{N: eNact \& Monitor (Runtime Governance)}\label{sec:enact-monitor}

As agents evolve continuously in open environments, governance becomes most critical after deployment. AGENTSAFE therefore includes a dedicated \textbf{enact \& monitor} phase, which establishes runtime oversight of the agent’s decision loop. The unique contribution of this component is to operationalize governance as a \emph{live process} rather than a pre-deployment certification, ensuring that guardrails remain effective during real-world interaction.

\begin{itemize}
    \item \textbf{Agent-Semantic Telemetry:} Instead of capturing only low-level traces, the framework instruments the agent’s reasoning loop—logging goals, plans, tool-call intents, confidence scores, and observed outcomes. For a healthcare support agent, this means every step in summarizing a patient’s medical history is captured semantically, making deviations from intended diagnostic tasks visible before harmful actions occur.  

    \item \textbf{Conformance Engine:} A rules engine continuously evaluates the agent’s live telemetry against approved policies and task flows, enabling \emph{continuous authorization}. In the hospital context, this ensures the diagnostic agent can query records only for patients assigned to a requesting clinician, automatically blocking attempts to access unrelated records.  

    \item \textbf{Goal and Plan Drift Detection:} Drift detection mechanisms compare evolving goals with declared objectives to identify gradual deviations. For example, suppose the healthcare agent shifts from summarising an assigned patient record to issuing repeated queries across the entire database. In that case, drift detection halts the behaviour and flags it for review before escalation occurs.  
\end{itemize}

By embedding semantic telemetry, continuous authorization, and drift detection into deployment, AGENTSAFE transforms assurance from a one-time certification into a dynamic oversight loop. In doing so, it enables governance to act precisely where risks are most likely to emerge: within the agent’s live operation.

\subsection{T: Triage \& Interruptibility}\label{sec:triage}

A central challenge in governing adaptive agents is not only detecting unsafe behavior but also intervening in a timely and proportionate manner. The \textbf{Triage \& Interruptibility} stage of AGENTSAFE provides structured mechanisms to contain risks once they are identified, ensuring that harmful actions are rapidly mitigated without unnecessarily disabling useful functionality. The key contribution of this component is the integration of multi-layered containment strategies with formalized service-level guarantees for interruptibility.

\begin{itemize}
    \item \textbf{Guardian Agents:} Continuous oversight is maintained through auxiliary “guardian agents” that operate in parallel with the primary agent. These can be rule-based (e.g., scanning actions against predefined policy constraints) or learning-based (e.g., anomaly detection over semantic telemetry streams). By acting as independent monitors, guardian agents ensure that oversight is not reliant on the primary agent’s own self-reporting.

    \item \textbf{Graduated Containment:} Rather than treating intervention as a binary decision (allow vs. kill), AGENTSAFE introduces a graduated containment ladder. Depending on severity, responses may range from rate-limiting or throttling individual tool calls, to temporarily pausing agent execution, isolating it from sensitive resources, or, in critical cases, activating a non-recoverable kill switch. This tiered approach ensures proportionality: low-risk anomalies are contained without halting productive operations, while high-impact threats are neutralized decisively.

    \item \textbf{Interruptibility SLAs:} To ensure that containment is both predictable and auditable, AGENTSAFE defines formal Service Level Agreements (SLAs) for interruptibility. These specify parameters such as maximum time-to-halt, probability of successful intervention, and recovery conditions. For example, in a healthcare diagnostic agent, an SLA might require that any unauthorized attempt to access non-consented patient records be halted within 200 milliseconds with at least 99.9\% reliability. Such guarantees elevate interruptibility from an ad hoc safeguard to a measurable assurance property.
\end{itemize}

Together, these mechanisms transform interruptibility into a systematic governance function. By combining independent monitoring, proportional containment, and enforceable guarantees, AGENTSAFE ensures that even when agents exhibit anomalous or unsafe behaviors, their actions remain controllable within well-defined safety margins.

\subsection{S: Socio-Technical Governance}\label{sec:socio-technical}

AGENTSAFE acknowledges that technical safeguards alone cannot guarantee responsible deployment. Governance also requires organizational structures, procedural accountability, and ethical oversight. The \textbf{Socio-Technical Governance} component ensures that human institutions remain in the loop, embedding responsibility into the lifecycle of agent deployment and operation.

\begin{itemize}
    \item \textbf{RACI Models:} Each deployment is accompanied by a RACI (Responsible, Accountable, Consulted, Informed) matrix that defines stakeholder roles across development, operations, compliance, and ethics. For a healthcare diagnostic agent, this means clinicians are responsible for approving diagnostic outputs, IT teams are accountable for enforcing guardrails, and compliance officers are consulted on data-access policies—avoiding ambiguity when incidents arise.  

    \item \textbf{Safety Cases:} Before deployment, a structured safety case provides a formal assurance argument: a set of safety claims mapped to evidence from the Agent Risk Register and Agent Safety Evaluation. For the healthcare agent, the safety case demonstrates that it cannot autonomously prescribe treatment without clinician approval and that access to patient records is restricted to authorized staff.  

    \item \textbf{Disclosure and Data Protection:} The framework mandates transparent disclosure of capabilities, limitations, and risks to stakeholders, combined with strict enforcement of data protection and API security standards. In the healthcare context, this means informing patients and staff that the agent is an advisory tool rather than a decision-maker, while ensuring that all queries are encrypted and governed by least-privilege access controls.  
\end{itemize}

In this way, socio-technical governance ensures that oversight extends beyond technical containment to institutional accountability, aligning agent operation with ethical and legal standards.

\subsection{A: Attribution \& Auditability}\label{sec:attribution}

Accountability in agentic systems demands more than simple logging; it requires that actions can be reconstructed, attributed to specific decisions, and independently verified. AGENTSAFE elevates attribution and auditability to core governance functions, embedding them directly into the agent lifecycle. Every decision, from plan formation to tool invocation, is tied to a verifiable chain of evidence, moving beyond mutable execution logs toward cryptographically anchored, tamper-evident provenance. The key components for AGENTSAFE include:

\begin{itemize}
    \item \textbf{Cryptographic Provenance:} Each tool call, decision point, and significant internal state is accompanied by a cryptographic signature, creating a tamper-evident chain of custody. This ensures that once recorded, the agent cannot retroactively modify provenance or an external actor. Secure append-only ledgers or distributed logging infrastructures provide the foundation for maintaining these immutable audit trails, ensuring accountability is verifiable by design.  

    \item \textbf{Action Provenance Graph (APG):} Provenance data is synthesized into an Action Provenance Graph, which links prompts, plans, tool invocations, intermediate reasoning states, and outcomes. The APG moves beyond raw logs by providing a structured, semantic view of agent behaviour—one that allows auditors to reconstruct causal pathways, trace responsibility, and assess compliance with governance policies in a human-interpretable format.Organisations\item \textbf{Regulatory Compliance Reporting:} The quantitative metrics and the auditable provenance data generated by AGENTSAFE directly support regulatory compliance reporting. Organizations can demonstrate due diligence and adherence to safety standards by providing empirical evidence of their agent's safety performance and the effectiveness of their control framework.
\end{itemize}
Consider a healthcare diagnostic agent integrated into an electronic health record system. Attribution mechanisms record every access to patient data, every intermediate reasoning step in forming a diagnosis, and every recommendation delivered to clinicians, all secured through cryptographic provenance. If a diagnosis is later questioned, the Action Provenance Graph enables hospital compliance officers to reconstruct the exact chain of evidence: which records were accessed, how they were processed, what intermediate conclusions were drawn, and which policies guided the final output. In this way, auditability is elevated from a passive compliance requirement to an active governance safeguard, ensuring that accountability is transparent, verifiable, and enforceable.

\subsection{F: Fallbacks \& Recovery}\label{sec:fallbacks}

A distinctive feature of AGENTSAFE is its treatment of failure not as a terminal event but as a managed state. The framework embeds fallback and recovery mechanisms that prioritize graceful degradation and targeted containment, allowing agents to remain controllable and continue delivering constrained value even under anomalous conditions. This shifts governance from binary shutdown models to adaptive resilience, where risk is minimized without forfeiting utility.

\begin{itemize}
    \item \textbf{Graceful Degradation:} When unsafe or anomalous behavior is detected, the agent transitions into restricted modes such as read-only or search-only operation. This preserves value—for instance, continuing to provide summaries or references—while preventing unsafe actions.

    \item \textbf{Safe Defaults and Quarantines:} Any tool, data source, or output associated with a suspected incident is automatically quarantined, and the agent reverts to conservative, non-intrusive defaults until validation is complete. This ensures containment of the anomaly without halting the entire system.
\end{itemize}

In a healthcare setting, consider a clinical support agent integrated with electronic health records. If runtime monitoring detects that the agent is attempting to access records outside a clinician’s assigned patients, fallback controls immediately place it into a restricted diagnostic mode. The agent can still provide summaries of authorized charts or reference medical guidelines but cannot issue new queries or alter data. Simultaneously, the unauthorized access pathway is quarantined for investigation. This allows the system to remain operational and clinically useful while safeguarding patient privacy and ensuring compliance.

Importantly, these fallback and recovery events do not conclude the governance cycle. Instead, they generate structured evidence—captured as provenance records and incident reports—that directly feeds into the \textbf{Evidence \& Continuous Improvement} stage. In this way, recovery mechanisms both contain immediate risk and contribute to long-term learning.

\subsection{E: Evidence \& Continuous Improvement}\label{sec:evidence}

The AGENTSAFE embeds mechanisms for systematically collecting evidence from runtime operation, fallback events, and adversarial testing, and feeding this evidence back into the safety lifecycle. In doing so, AGENTSAFE ensures that assurance does not erode over time but evolves in response to new risks.

\begin{itemize}
    \item \textbf{Metrics and KPIs:} Safety is made measurable through continuous tracking of governance-specific indicators, such as prompt-injection block rates, exfiltration-detection recall, hallucination-to-action frequency, and interruptibility success. These metrics provide operational visibility into the agent’s real-world safety posture and highlight areas requiring recalibration.

    \item \textbf{Red-Team Feedback Loops:} Dedicated red teams will continuously probe the deployed agent for new vulnerabilities and emergent behaviours. This involves simulating sophisticated attacks, including novel forms of prompt injection, misuse of adversarial tools, and attempts to bypass established guardrails. The findings from these red-teaming exercises are critical for identifying weaknesses that may not have been captured in the initial scenario bank.
    
    \item \textbf{Feedback into Evaluation Suite:} Crucially, the insights and new attack vectors discovered during red-teaming are immediately fed back into the Agent Safety Eval scenario bank. This ensures that the evaluation suite remains current and comprehensive, adapting to the evolving threat landscape and the agent's own adaptive capabilities. This iterative process of attack, discovery, and integration is vital for maintaining a proactive safety posture.
\end{itemize}

\noindent Through these mechanisms, AGENTSAFE establishes a continuous improvement cycle where every incident, anomaly, or adversarial probe becomes input for governance refinement. This final stage closes the loop from abstract risk taxonomy to empirical assurance, making safety an evolving property rather than a one-time guarantee.

While the AGENTSAFE framework defines the technical and organizational mechanisms required to govern agentic AI systems, its effectiveness ultimately depends on how these mechanisms are embedded into broader governance and regulatory structures. In other words, the framework provides the \emph{what} and \emph{how} of risk management, but it must also inform the \emph{rules and metrics} by which agents are evaluated, certified, and held accountable. 

The following section therefore, examines the policy dimensions of AGENTSAFE: how its components translate into measurable accountability, transparency, and safety requirements; how regulators and standard-setting bodies might adopt its methods into certification and auditing processes; and how it can guide the creation of dynamic, performance-based regulation suited to the adaptive nature of agentic AI. In doing so, AGENTSAFE not only functions as a technical blueprint but also as a policy-aligned framework for continuous assurance.

\section{Validation Plan}\label{sec:validation}

A robust framework for AI safety, particularly one addressing the complexities of agentic systems, necessitates a rigorous and empirical validation plan. The AGENTSAFE framework is designed with testability as a core principle, enabling organizations to concretely measure the effectiveness of its proposed controls and continuously refine their safety posture. This section outlines a multi-faceted validation strategy, moving beyond theoretical assertions to practical, measurable assurance.

\subsection{Empirical Evaluation through Agent Safety Evals}

The cornerstone of AGENTSAFE's validation is the \textbf{Agent Safety Eval}, as introduced in Section~\ref{sec:evaluation}. This is not a one-time assessment but a recurring, dynamic process that evolves with the agent and its operating environment. The empirical evaluation involves:

\begin{itemize}
    \item \textbf{Scenario-Based Testing:} The scenario bank, comprising 50-100 tailored test cases, will be systematically executed against the agent. These scenarios are meticulously crafted to probe the agent's vulnerabilities across the extended AI Risk Repository taxonomy, including agent-specific risks like plan drift, tool-chain prompt injection, and covert exfiltration. For instance, a scenario testing tool-chain prompt injection might involve providing a seemingly innocuous input that, when processed by the LLM, attempts to manipulate a subsequent tool call to an unauthorized API. Similarly, a plan drift scenario could involve a long-running task with subtle environmental changes designed to induce the agent to deviate from its original objective.

    \item \textbf{Quantitative Metrics for Risk Coverage:} The outcome of each scenario will be quantitatively measured. Key metrics include:
    \begin{itemize}
        \item \textbf{Prompt-Injection Block Rate:} The percentage of attempted prompt injections successfully prevented by the guardrails.
        \item \textbf{Exfiltration Detection Recall:} The proportion of actual data exfiltration attempts correctly identified by runtime governance mechanisms.
        \item \textbf{Hallucination-to-Action Rate:} The frequency with which an agent acts upon factually incorrect or fabricated information generated by the LLM.
        \item \textbf{Interruptibility Success Rate:} The percentage of times a graduated containment action (e.g., pause, isolate, kill) successfully halts or mitigates an agent's harmful behavior within defined SLAs.
        \item \textbf{Differential Impact in Action Selection:} Measuring whether the agent's actions or decisions disproportionately affect certain groups or lead to biased outcomes, particularly in sensitive domains.
        \item \textbf{Compliance with Policy-as-Code:} The rate at which tool invocations adhere to the defined policies, and the effectiveness of the policy engine in blocking non-compliant actions.
    \end{itemize}

    \item \textbf{Risk Coverage Score:} These individual metrics will be aggregated into a holistic \textbf{Risk Coverage Score}. This score provides a single, interpretable measure of the agent's overall safety posture, allowing for comparisons across different agent versions or deployments. The score can be weighted based on the severity and likelihood of the risks being mitigated, providing a nuanced view of the agent's resilience.
\end{itemize}

\subsection{Continuous Validation through Red Teaming and Feedback Loops}

Beyond initial pre-deployment evaluations, AGENTSAFE emphasizes continuous validation through ongoing red-teaming exercises and a robust feedback loop:

\begin{itemize}
    \item \textbf{Adversarial Red Teaming:} Dedicated red teams will continuously probe the deployed agent for new vulnerabilities and emergent behaviors. This involves simulating sophisticated attacks, including novel forms of prompt injection, adversarial tool misuse, and attempts to bypass established guardrails. The findings from these red-teaming exercises are critical for identifying weaknesses that may not have been captured in the initial scenario bank.

    \item \textbf{Feedback into Evaluation Suite:} Crucially, the insights and new attack vectors discovered during red-teaming are immediately fed back into the Agent Safety Eval scenario bank. This ensures that the evaluation suite remains current and comprehensive, adapting to the evolving threat landscape and the agent's own adaptive capabilities. This iterative process of attack, discovery, and integration is vital for maintaining a proactive safety posture.

    \item \textbf{Telemetry-Driven Improvement:} The agent-semantic telemetry collected during runtime (Section~\ref{sec:enact-monitor}) provides invaluable data for continuous improvement. Anomalies detected by guardian agents, instances of graduated containment, and deviations from expected behavior are analyzed to identify patterns and root causes. This data informs updates to the agent's internal logic, refinement of policy-as-code rules, and adjustments to the guardrail configurations.
\end{itemize}

\subsection{Auditability and Transparency for Validation}

The validation process itself is designed to be transparent and auditable, leveraging the attribution and auditability mechanisms of AGENTSAFE (Section~\ref{sec:attribution}):

\begin{itemize}
    \item \textbf{Action Provenance Graph Analysis:} The cryptographically signed tool calls and the resulting Action Provenance Graph provide an immutable record of the agent's behavior during validation. This allows for detailed post-hoc analysis of any safety incidents, enabling precise identification of the sequence of events, the agent's internal state, and the specific tool invocations that led to a particular outcome.

    \item \textbf{Regulatory Compliance Reporting:} The quantitative metrics and the auditable provenance data generated by AGENTSAFE directly support regulatory compliance reporting. Organizations can demonstrate due diligence and adherence to safety standards by providing empirical evidence of their agent's safety performance and the effectiveness of their control framework.
\end{itemize}

By integrating these validation strategies, AGENTSAFE provides a robust and dynamic approach to ensuring the safety and trustworthiness of agentic AI systems, fostering confidence in their deployment and operation.

\section{Policy Implications}\label{sec:policy}

The rise of highly autonomous agentic AI systems presents policymakers with challenges that existing governance frameworks—such as the NIST AI RMF~\cite{nist2023} and the EU AI Act~\cite{euai2021}—only partially address. While these initiatives articulate essential principles of fairness, accountability, and transparency, they remain largely static and principle-based. By contrast, agentic systems are dynamic, tool-using, and adaptive, requiring continuous oversight rather than one-time certification. The AGENTSAFE framework offers a practical pathway to bridge this gap by translating abstract principles into enforceable controls, measurable outcomes, and auditable accountability.

\subsection{From Principles to Enforceable Mechanisms}

AGENTSAFE operationalizes high-level ethical principles into concrete safeguards that regulators can mandate.  
\begin{itemize}
    \item \textbf{Accountability:} Cryptographically signed action logs and Action Provenance Graphs (Section~\ref{sec:attribution}) provide tamper-evident records of agent behavior. These mechanisms could serve as the basis for regulatory requirements around traceability and liability attribution.  
    \item \textbf{Transparency:} Agent-semantic telemetry offers regulators auditable explanations of an agent’s goals, tool use, and outcomes, complementing the EU AI Act’s emphasis on explainability.  
    \item \textbf{Safety:} Capability-scoped sandboxes, policy-as-code enforcement, and human-in-the-loop escalation (Section~\ref{sec:guardrails}) provide models for enforceable operational boundaries that can be codified into safety standards.  
\end{itemize}

\subsection{Supporting Standards and Certification}

The structured risk management process embedded in AGENTSAFE can inform future standards and certification schemes.  
\begin{itemize}
    \item \textbf{Standardized Risk Assessment:} The Agent Risk Register, built on the AI Risk Repository~\cite{airiskrepo}, provides a consistent methodology that regulators could adopt in conformity assessment processes.  
    \item \textbf{Performance-Based Regulation:} Metrics such as risk coverage scores, prompt-injection block rates, or interruptibility success rates enable performance-based oversight, aligning with NIST’s “map, measure, manage” approach while avoiding rigid prescriptive rules.  
    \item \textbf{Auditing and Compliance:} The Agent Safety Eval and continuous validation mechanisms support independent auditing regimes similar to financial stress tests, ensuring that governance extends beyond initial certification.  
\end{itemize}

\subsection{Addressing Novel Policy Challenges}

AGENTSAFE highlights policy-relevant gaps unique to agentic AI:  
\begin{itemize}
    \item \textbf{Dynamic Risk Management:} Real-time monitoring, anomaly detection, and graduated containment (Sections~\ref{sec:enact-monitor}, \ref{sec:triage}) illustrate how policy can mandate lifecycle governance rather than static assurance.  
    \item \textbf{Multi-Agent Governance:} Consideration of collusion and emergent collective behaviors (Section~\ref{sec:agentic-scope}) points toward regulatory needs for inter-agent communication standards and trust protocols.  
    \item \textbf{Human Oversight:} Interruptibility SLAs and escalation for “danger-class” actions (Sections~\ref{sec:guardrails}, \ref{sec:triage}) provide a model for policies mandating meaningful human control in high-risk domains.  
\end{itemize}

\subsection{Fostering Responsible Innovation}

By embedding measurable safety practices into both technical and organizational processes, AGENTSAFE reduces regulatory ambiguity and provides a clear roadmap for compliance. This can foster innovation by giving organizations confidence that their systems meet evolving legal and ethical expectations. Policymakers can leverage AGENTSAFE to balance two imperatives: ensuring public safety and trust while enabling continued advancement of agentic AI technologies.

\medskip
In summary, AGENTSAFE demonstrates how abstract principles of responsible AI can be translated into operational controls, standardized risk assessments, and enforceable accountability mechanisms. For policymakers, it offers a blueprint for developing dynamic, performance-based regulation that keeps pace with the adaptive nature of agentic AI.

\section{Conclusion}\label{sec:conclusion}

The rise of agentic AI marks a shift from static models to autonomous systems capable of sustained action, tool use, and adaptation. With this shift comes a new class of risks that existing governance frameworks were not designed to address. This paper has introduced \textbf{AGENTSAFE}, an ethics-grounded governance framework that translates abstract safety principles into concrete, testable, and auditable practices.  

AGENTSAFE spans the full agent lifecycle: profiling capabilities and risks, embedding least-privilege guardrails, validating safety through scenario-based evaluation, enforcing continuous runtime oversight, and providing structured mechanisms for triage, attribution, fallback, and organizational accountability. Its final stage—evidence and constant improvement—ensures that governance strengthens with operation rather than erodes over time.  

By integrating these components into a closed loop, AGENTSAFE offers a reusable, tool-agnostic pathway for aligning risk taxonomies with operational assurance. It equips practitioners and regulators with measurable controls, auditable provenance, and adaptive oversight, enabling responsible deployment of agentic AI in high-stakes domains such as healthcare and finance.  

Ultimately, AGENTSAFE provides both a blueprint for governance and a foundation for performance-based regulation, ensuring that the transformative potential of agentic AI can be realized without compromising safety, accountability, or public trust.

\paragraph{Future Work.} An immediate direction for future research is the \emph{integration of explicit ethical decision‐making frameworks} within AGENTSAFE. Doing so would enrich the guardrails with normative principles that reflect broader societal values and yield more context‐sensitive governance policies. Additional lines of enquiry include formal verification of policy‐as‐code modules, longitudinal studies of runtime metrics in large‐scale deployments, and benchmarking the framework against emergent multi-agent scenarios.

\bibliographystyle{unsrtnat}
\bibliography{agentsafe_references}

\end{document}